\documentstyle [11pt]{article}
\def\be{\begin{equation}}
\def\ee{\end{equation}}
\begin{document}
\baselineskip=14pt
\title{The Case for a Gravitational de Sitter Gauge Theory}
\author{R. Aldrovandi and J. G. Pereira \\
Instituto de F\'{\i}sica Te\'orica,
Universidade Estadual Paulista\\
Rua Pamplona 145, 01405-900\, S\~ao Paulo\, SP -- Brazil}
\date{}
\maketitle
{\hskip 4.0cm {\sc In homage to Paulo Leal Ferreira on}}

{\hskip 4.8cm {\sc the occasion of his 70th birthday}}

\vspace{0.5 cm}
\begin{abstract}
With the exception of gravitation, the known fundamental interactions of 
Nature are mediated by gauge fields. A comparison of the candidate groups 
for a gauge theory possibly describing gravitation favours the Poincar\'e 
group as the obvious choice. This theory gives Einstein's equations in a 
particular case, and Newton's law in the static non-relativistic limit, 
being seemingly sound at the classical level. But it comes out that it is 
not quantizable. The usual procedure of adding counterterms to make it a 
consistent and renormalizable theory leads to two possible theories, one 
for each of the two de Sitter groups, SO(4,1) and SO(3,2). The consequences 
of changing from the Poincar\'e to the de Sitter group, as well as the 
positive aspects, perspectives and drawbacks of the resulting theory are 
discussed.
\end{abstract}

\section{Introduction}

General Relativity is the widely accepted theory of gravitation. Besides 
its consistency and beauty, it has accumulated an impressive amount of 
experimental successes, which seems to establish its validity beyond any 
possibility of doubt. It was submitted to a very heavy attack some years 
ago~\cite{1}, to which followed an equally passionate defense~\cite{2}. 
But the debate was, curiously enough, restricted to the soviet community. 
Independently of the hard core of the subject, on which we shall not take 
position here, it is undeniable that such controversies are highly to be 
praised. No theory can be accepted as eternal, or incapable of improvement 
and any honest, good-willed attack is always healthy. A consequence of the 
polemics has been a reapraisal of the accepted theory by its very supporters, 
which was a very positive point. Some weak points, well known to experts, 
have become of widespread knowledge. Even true believers of a theory accept 
that alternative models help its understanding and provide references for 
experimental improvements. Some alternative theories of the past 
(Brans-Dicke, for example) have played such a "sparring" role.
 
Are there additional reasons to look for alternative theories? From the 
experimental point of view, there is none. General Relativity has an 
impressive record of experimental successes. We could estimate that it 
has been verified, in those cases in which it was possible, to around 0.1\%. 
This is far less precise than the analogous score for some other 
interactions, but experiments are exceedingly difficult. From the 
theoretical point of view, there is no imperative, compeling reason. 
There are a few reasons nevertheless.
 
Some theoretical {\it defects} of General Relativity sprang out in 
the controversy alluded to. To start with, let us talk of two of them. 
First, in order to have things well-defined, even the gravitational 
field should be asymptotically vanishing. This means that, in particular, 
the space sector of spacetime should be asymptotically flat and would 
lead to trouble with one of the two favoured universe models of Cosmology: 
it would be consistent with the open Friedmann universe, but not with 
the closed Friedmann universe. A second difficulty is that General 
Relativity is not really a field theory. The argument runs as follows: 
when talking about a field, we must be able to say where it is. The 
usual means for that is to calculate its energy density: the field is 
there where the energy density is different from zero. General 
Relativity attributes no well-defined energy density to the gravitational 
field. It is true however that an analogous difficulty is present in 
gauge theories. There, also the "charge" density of the gauge field is 
ill-defined. Thus, for example, the color density of the gluon field 
has no physical meaning. Energy (the energy-momentum tensor) is the 
"charge" for  gravitation and, just as for the gauge charges in general, 
only the total charge of gauge field plus source fields has a covariant 
meaning \cite{3}. There is also the possibility of interpreting General 
Relativity as a metric field supperposed to a Minkowski background. 
This would not change the problem of the energy, but would make it still 
more similar to the gauge charge problem. By the way, a frequent 
argument against General Relativity is that the energy-momentum tensor 
is the Noether current for translations, and that the remaining local 
symmetries of spacetime (that is, the Lorentz transformations) are not 
contemplated. 

There are a number of further arguments against General Relativity. A 
very common one, which is weak to the point of being almost wrong, would 
run more or less as follows: the desired goal is an unified theory, and 
this would require similar theories. Well, amongst the 4 interactions 
nowadays taken as fundamental, the electromagnetic and weak interactions 
are described by the Weinberg-Salam gauge model, and the strong 
interactions are described by the SU(3) gauge chromodynamics. And 
these theories are also very succesful! Only gravitation stays apart 
from the 3 gauge-ruled interactions. It should then, {\it unification 
oblige}, have also a gauge formulation. Now, the truth is that 
unification can be conceived even if gravitation remains {\it different}. 
In this case, there are two possibilities: (i) it remains different, 
but it is quantizable, or (ii) it is so different that it is not 
quantizable. A favorite possibility of the last type is "induced 
gravitation" in the sense of Zeldovich and Sakharov~\cite{6}, which 
conceives gravitation as a kind of elastic property of spacetime, 
coming from the vacua of the other interactions. It would be necessary, 
in order to support this idea, to show that such vacua do induce a 
curvature on spacetime. Unfortunately, despite gigantic efforts, this 
has not been shown up to now. This vision of gravitation as an 
effective interaction would allow it to remain different from the 
other and, furthermore, to remain essentially classical.

The very success of General Relativity, on the other hand, teaches 
two important things to those in search of alternative models: (i) 
of course, the alternative model should give the same well-verified 
results, and (ii) gravitation does exhibit a privileged relationship 
to spacetime.
All this has led to two main kinds of alternative models. The first 
one is the higher-order curvature Lagrangians, like ${\cal L}=R + R^2$. 
The differences with respect to the Lagrangian $R$ are known of 
old~\cite{7}, but are not observable with present-day resources. 
Though improving renormalization, these Lagrangians require 
successive addition of terms (called counterterms) to account for 
divergences at higher perturbation orders. This means that we 
should have actually something like ${\cal L}=R + R^2 + R^3 + R^4 + 
\dots \;$. And here we notice that we work with a beloved prejudice, 
whose main justification lies in simplicity: we suppose the Lagrangian 
to be a polynomial in the field and their derivatives. Why not things 
like ${\cal L} = 1/(1-R)$, for example? Non-polynomial Lagrangians 
have been fashionable in the seventies, but seem to have been 
abandoned by now. The second kind of alternative models is super-
gravity~\cite{8}, which adds a particle of opposite statistics to 
each known particle, as well as improves renormalizability for the 
lower order perturbative graphs, but fails at higher orders. 
There is still another point of view: since gravitation is different, 
let us quantize it differently! This is the banner of the so called 
"Quantum Gravity" scheme. Recently, Ashtekhar proposed another 
approach~\cite{9}, a version of the Hamiltonian formalism in which 
the gravitational variables appear as gauge variables, with the 
advantage that some of the gauge constraints are automatically 
satisfied.

We go now to the main point of our paper. There is another frequent 
argument, which is wrong but is relevant because it calls attention 
to an important point: the incompatibility of General Relativity 
with quantum requirements. It is attributed to Bohr and Rosenfeld 
the statement: every field must be quantized, since otherwise it 
would be possible to violate the uncertainty principle. This is a 
rather loose version of what is really said in the famous Bohr-
Rosenfeld paper~\cite{4}. As Rosenfeld himself pointed out~\cite{5}, 
the arguments, there applied to the electromagnetic field, do not 
apply to gravitation. Whether gravitation is to be quantized or not, 
the answer is to be given by experiment.

The fact remains, however, that General Relativity is not (at least 
perturbatively) renormalizable. Supported on their natural affinity 
to renormalization, a large number of gauge models for gravitation 
have been proposed. Our intention here will be to present the case 
for one of them, the Gravitational de Sitter Gauge Model. Because 
the subject is very wide \cite{10}, we shall be a bit na\"{\i}ve 
and adopt a rather assertive style, even at the risk of seeming 
dogmatic. In broad brushstrokes, the case is summed up in the 
following points: (i) gravitation is deeply related to spacetime 
itself, much more so than other nowadays known interactions; (ii) 
gauge theories describe suitably the other interactions and, 
despite the above discussion about the weakness of this argument, 
it seems natural for us to look for alternatives inspired by the 
gauge scheme; (iii) the natural group to be considered is the 
Poincar\'e group; (iv) a gauge theory for the Poincar\'e group is 
plagued with a deadly illness: it has no action functional; (v) if 
"quantized" in a way that dispenses with the action functional, it 
has no well-defined vertices, a problem that can be solved by a 
method inspired by renormalization theory, that is, by the addition 
of counterterms; (vi) once this is done, the resulting theory is 
non-renormalizable at first, but addition of new counterterms turn 
it into a renormalizable theory; (vii) the resulting model, once 
the counterterms have been added, is a gauge theory for one of the 
de Sitter groups. The gauge theory for a de Sitter group appears 
consequently as a smoothed, renormalizable Poincar\'e gauge theory.

As the crux of the problem lies in the question of renormalizability, 
we start by a brief discussion of the subject in section~2. Dimensional 
considerations lead to one of the main arguments favoring de Sitter 
gauge theories: it is very difficult to conceive a renormalizable theory 
with the energy-momentum as source current. With the exception of 
gravitation, the known fundamental interactions of Nature are mediated 
by gauge fields. The general, formal characteristics of gauge theories 
are summed up in section~3, followed by a comparison of the candidate 
groups for a gauge gravitation theory. The Poincar\'e group comes out 
as the obvious choice. We consequently analyse the gauge theory for the 
Poincar\'e group in the following section, together with a short 
discussion of the bundle of linear frames which appears as the geometric 
background. The theory gives Einstein's equations in a particular case, 
and Newton's law in the static non-relativistic limit, being seemingly 
sound at the classical level. But it comes out that it is not quantizable. 
As described in section~5, the procedure of adding counterterms to make 
it into a consistent and renormalizable theory leads to two possible 
theories, one for each of the two de Sitter groups, SO(4,1) and SO(3,2). 
The consequences of changing from the Poincar\'e to the de Sitter group, 
as well as the positive aspects, perspectives and drawbacks of the 
resulting theory, are outlined in the final section.

\section{Renormalizability}

A theory which does not bow to the renormalizability requirement is 
unacceptable from the quantum point of view: it will attribute infinite 
values to finite quantities. There seems to be theories which are 
renormalizable even if not perturbatively so, but this is too involved 
a subject to be considered here. We speak here of perturbative, order 
by order renormalizability. This is a very restrictive condition. 
Amongst all the polynomial models, there are only three types of 
renormalizable theories: \\
(1) scalar theories with interaction term of type $\lambda \phi^4$; \\
(2) scalar-fermion interactions of Yukawa type: $g \bar{\Psi} \Psi \phi$ 
for scalars and  $g \bar{\Psi} \gamma^5 \Psi \phi$  for pseudo-scalars; \\
(3) minimal coupling as given by gauge theories for reasonable groups 
(like SU(N) and SO(N)). \\
An essential characteristic coming out from the detailed examination of 
the problem is embodied in a simple rule-of-thumb: in order that the 
theory be renormalizable, the coupling constants must be non-dimensional. 

Elementary dimensional analysis is of help here. One uses a system of 
units in which $\hbar=c=1$, that is, they are non-dimensional: $[\hbar] = 
[c] = 0$. One counts the dimension in terms of the mass: $[m] = 1$. 
This means that some usual quantities have dimensions like 
$[\partial_{\mu}] = [E] = [T^{-1}] = [L^{-1}] = 1$. Comparing the 
different terms in the free Lagrangians, one finds that the main 
fields have the following dimensions: bosons, $[\phi] = [A_\mu] = 1$; 
fermions, $[\Psi] = [\bar{\Psi}] = 3/2$; field strengths of gauge 
theories, $[F_{\mu \nu}] = 2$. A very general result is that all 
Noether source currents have $[J] = 3$. 

A disturbing fact appears in General Relativity, in which the Noether 
source current has dimension $[T_{\mu \nu}] = [E/V] = 4$. This is clearly 
related to the fact that the usual transformation generators are 
dimensionless, except those of translations, which have $[\partial_{\mu}] 
= 1$. It is true that also the dimensions of the basic fields are 
anomalous in General Relativity: $[R] = 2$ is a correct field strength, 
but the metric $g_{\mu \nu}$ has $[g_{\mu \nu}] = 0$! The final trouble 
comes really from the energy-momentum "irregularity": the coupling 
constant $k$ is seen, from $[R_{\mu \nu}] \sim k[T_{\mu \nu}]$, to 
have dimension $[k] = - 2$. The source current, being the Noether 
current associated to translations, enforces a non-vanishing dimension 
for the coupling constant $k$ in General Relativity. It is anyhow hard 
to imagine a renormalizable theory with the energy-momentum as source 
current, which is itself a non-renormalizable tensor. In the perturbative 
series, each term contains Feynman integrals in the momenta and must 
be dimensionless. Each vertex produces a factor $f$, here with $[f] = 
[L^2]$. In order to compensate this dimension, some factor $p^2$ turns 
up in the integrand and the terms become more and more divergent as the 
order of perturbation increases~\cite{11}. The measure of divergence is 
given by the superficial degree of divergence $\hat{w}_v$ for each 
vertex, which is given by $\hat{w}_v = \delta_v + (3/2) f_v + b_v$, 
where $\delta_v$ is the number of derivatives in the internal lines 
incident at the vertex $v$, $f_v$ is the number of incident fermion 
lines, and $b_v$ is the number of boson lines. If, for all vertices, 
$\hat{w}_v > 4$, the theory is non-renormalizable; if, for all vertices, 
$\hat{w}_v \leq 4$, then the theory is renormalizable; if, for all 
vertices, $\hat{w}_v < 4$, the theory is super-renormalizable. It turns 
out that renormalizability may be checked by inspection of the 
interaction terms in the Lagrangian. In last resort, what happens is 
that the constants in the Lagrangian are corrected order by order. 
One says then that the infinities are absorbed in the constants. 
These constants include the wavefunction normalizations, the masses, 
and the coupling constants. If they are enough to absorb all the 
infinities, the theory is renormalizable. Actually, if we find that 
a given theory is non-renormalizable, it can eventually be "repaired": 
it may happen that it becomes renormalizable once one or more terms 
("counterterms") are added to the primitive Lagrangian. The theory is 
renormalizable when the number of necessary counterterms is finite. 
The physical starting Lagrangian must contain all the necessary 
counterterms. 

Let us look at the simplest example, the electrodynamics of mesons 
(such as $\pi^+$ and $\pi^-$ mesons). The free Lagrangian density 
would be
$$
{\cal L} = - \partial_{\mu} \phi^* \partial^{\mu} \phi + m \phi \phi^* -  
\frac{1}{4} F^{\mu \nu} F_{\mu \nu} \, .
$$ 
The interaction is added through the gauge prescription, by which 
ordinary derivatives are replaced by covariant ones: $\partial^\mu 
\rightarrow  \partial^\mu - i e A^\mu$. The Lagrangian density becomes
$$
{\cal L} = - [\partial_\mu + i e A_\mu] \phi^* [\partial^\mu - i e A^\mu] 
\phi + m \phi \phi^* - \frac{1}{4} F^{\mu \nu} F_{\mu \nu} \, .
$$ 
We then proceed to obtain the quantized, renormalized theory. A strange 
thing happens then: a theory as above is not consistent. In order to 
renormalize the graphs with four external meson legs and internal photon 
loops, it is necessary to add a counterterm $+\lambda |\phi \phi^*|^2$ 
to the above Lagrangian. This is why one always starts with the meson 
Lagrangian with the term $\lambda \phi^4$: one knows it will be 
necessary. It is a fascinating point that, in order to interact 
correctly through the exchange of photons, the mesons are forced to 
interact also between themselves, and in that particular prescribed way.

What happens in General Relativity is that new counterterms must be 
added at each order in perturbation, so that actually an infinite 
number of counterterms is required for the whole perturbative series. 
And, we repeat, this trouble comes from the "irregular" dimension of 
the energy-momentum tensor. A fact which, by the way, suggests that 
any theory with this tensor as a source will have the same kind of 
problem. Notice that some people argue that non-renormalizability 
may ultimately be a good thing for General Relativity. The requirements 
of field theory supposes Minkowski space at arbitrarily short distances, 
but there could be a natural cut-off given by the Planck scale~\cite{12}.

Notice that no interaction mediated by a vector meson is renormalizable 
{\it unless} the meson is a gauge boson. In this case, the gauge 
symmetry may be spontaneously broken, so as to endow the intermediate 
bosons with mass while preserving renormalizability. All that said, 
it seems reasonable, when looking for alternative theories for 
gravitation, to go after a gauge model, knowing nevertheless that 
that model should have a privileged relationship to spacetime. 

\section{Gauge theories}

What finally makes gauge theories~\cite{13} so especial? They have 
some really nice properties: \\
(i) they embody an automatic prescription for taking symmetries into 
account. But attention: such symmetries are particle-classifying, 
the elementary particles are placed in multiplets of the symmetry group; \\
(ii)  they have a natural affinity with renormalization; as said above, 
this is a real ace! \\
(iii) they have a very rigid structure, basically geometrical in 
character; they possess much more symmetry than that included in 
the gauge group: duality symmetry, conformal symmetry, BRST symmetry, etc.

Given a classifying group $G$ (a group in whose multiplets the 
elementary particles can be coherently accommodated) with Lie 
algebra $G^{\prime}$ generated by operators ${J_a}$ satisfying 
$[J_a, J_b] = f^{c}{}_{ab} J_c$, the gauge potential will have 
the form  $A_\mu = J_a A^{a}{}_{\mu}$, and the field strength will 
be $F_{\mu \nu} = J_a F^{a}{}_{\mu \nu} = J_a(\partial_\mu A^{a}{}_{\nu} - 
\partial_\nu A^{a}{}_{\mu} + f^{a}{}_{bc} A^{b}{}_{\mu} A^{c}{}_{\nu})$. 
We know perfectly the geometry behind it: $A$ is a connection and $F$ 
is its curvature, satisfying automatically the Bianchi identity
$$ 
\left(\delta^{a}{}_{c} \partial_\mu + f^{a}{}_{bc} A^{b}{}_{\mu} \right) 
\tilde{F}^{c \mu \nu} = 0 \, ,
$$
where $\tilde{F}^{c}{}_{\mu \nu} = \frac{1}{2} \epsilon_{\mu \nu \rho 
\sigma} {F}^{c \rho \sigma}$ is the dual of $F$. A change of gauge is 
a transformation  $A \rightarrow A^{\prime} = U(A + d)U^{-1}$, leading 
to $F^{\prime} = dA^{\prime} + A^{\prime}  \wedge A^{\prime} = U(dA + A 
\wedge A )U^{-1} = U F U^{-1}$. The field strength $F$, whose components 
are the measurable quantities, is covariant under gauge transformations. 
This means that they have, as they should, a covariant physical meaning. 
The basic dynamics is given by the Yang-Mills equation 
$$
\left(\delta^{a}{}_{c} \partial_\mu + f^{a}{}_{bc} A^{b}{}_{\mu} \right) 
{F}^{c \mu \nu} = J^{a \nu} \, ,
$$
where $J^{a \nu}$ is the source Noether current. Given the structure 
constants $f^{a}{}_{bc}$ of the group, one can always write directly the 
field  equations, one for each group generator. We profit here to make it 
clear what we mean by a gauge theory: it is a theory whose basic dynamics 
is governed by the Yang-Mills field equations.

In the sourceless case, $J^{a \nu} = 0$, the Yang-Mills equation is just 
the expression of the Bianchi identity written for the dual of $F$. This 
is the duality symmetry. Both equations are invariant if we change the 
space time metric $g_{\mu \nu}$ by multiplying it by a function: 
$g_{\mu \nu} \rightarrow  f(x) g_{\mu \nu}$. This is the conformal 
symmetry, deeply related to the renormalizability of the theory. 
This is of course no place for a detailed exposition of gauge theories. 
We only quote these items to give an idea of how rigid they are, so 
much so that you cannot change anything without breaking the whole 
structure and losing their good properties. The background structure~\cite{li}  
is well-known: its a principal fiber bundle, with spacetime as the 
base space and the gauge group as the fiber. The bundle space is 
locally a direct product of spacetime by the group. The connection 
$A$ takes vector fields on the bundle space into the Lie algebra of 
the gauge group. It represents the field mediating the interaction 
between source fields. 

Each source field is in an associated bundle, structure similar to 
the principal bundles but with a representation (a multiplet to which 
the source field belongs) replacing the group. There are in principle 
infinite connections on each bundle, amongst which the Yang-Mills 
equation, with suitable boundary conditions, chooses one. This fixes 
the gauge field of the problem under consideration.  

An important question is related to another dear prejudice: 
universality. Though experimental evidence is still lacking, there 
is a widespread belief that gravitation concerns all  elementary 
particles. There would be no particle that does not  feel  gravitation. 
The fact that the Poincar\'e group classifies all the elementary 
particles is not enough to ensure such property. Given a gauge model, 
particles insensible to the field are classified in the singlet 
representations, which are one-dimensional and whose dynamics will 
automatically "vanish". Scalar fields are singlets of the Lorentz 
group, though not of the translation group. This is one reason for 
the importance of translations --- they would "explain" universality. 
But there is another origin for universality, holding for any group 
acting on spacetime: the so called kinematic representations~\cite{27}. 
All these groups have a representation in terms of vector fields 
(that is, derivatives) on spacetime, which act through the arguments 
of the wavefunctions. This is true for translations, rotations, 
boosts, conformal transformations, and dilatations. 
Given any Lie group, it is an easy task to build up a  formal gauge 
theory by writing down the corresponding Yang-Mills equations. The 
presence of these kinematic representations, however, changes the 
scheme a lot. And they will, we repeat, be at work for any group 
acting on spacetime. 

There will be another, deep problem concerning gravitation and gauge 
theories. The latter have mediating fields of spin $1$. The interaction 
will consequently reverse sign when one of the interacting particles 
is changed into its antiparticle \cite{15}. This affects another 
beloved prejudice: that matter and antimatter have the same, 
attractive, gravitational interaction. We shall see later how this 
problem may come to be circumvented. 

The arguments listed above support the idea that, if we are to look 
for an alternative theory of the gauge type~\cite{14}, a group 
classifying the elementary particles must be involved, which should 
be intimately related to spacetime itself. Let us then briefly 
review the main groups acting on spacetime:\\
(i) the conformal group: it is deeply related to the causality 
structure~\cite{16}, as it contains the transformations preserving 
the light cones; it should be somehow broken, as its representations 
can only accommodate particles of vanishing masses; in other words, 
it does not really classify the known elementary particles;\\
(ii) the de Sitter groups: we will come back to them later;\\
(iii) the Lorentz group: it has been studied by Yang~\cite{17}, 
Camenzind~\cite{18}, Carmeli~\cite{19}, and many others; it does not
 really classify the elementary particles --- it would account for 
 spin but leaves momentum out of the game;\\
(iv) Poincar\'e group \cite{20}: it classifies the elementary particles, 
giving them both spin and momentum; it has a very clear relationship 
with spacetime, and is the obvious natural candidate; it is however 
a non-semisimple group and we shall see that this leads to a lot of 
trouble;\\
(v) the translation group: in certain aspects, it has been 
used~\cite{21} to rephrase General Relativity, but it does not 
classify the particles --- it takes only momentum into account. 

Thus, the Poincar\'e group, which is both the classifying group in 
what concerns spacetime and the basic local group of Physics, appears 
as the natural candidate for a gauge model for gravitation. Indeed, 
it has been the most studied group~\cite{22}, and we shall use it as 
the starting point of our analysis. It will have apparently unsolvable 
problems with quantization. Actually the best argument for the de Sitter 
group is that it comes out of the whole analysis as the "corrected" 
Poincar\'e group, in a sense to be made clear in the following.  

\section{The Poincar\'e group}

We have been saying that gauge theories have as background a principal 
fiber bundle, with gauge group as the fiber and spacetime as the base 
space. This is actually the way of doing geometry in (rather) modern 
language~\cite{23}. And we are, repeating again, looking for a gauge 
theory somehow linked to spacetime, much more so than the usual gauge 
models. Now, there is an important fact. Every differentiable manifold 
$M$ (here we are thinking of spacetime, of course) has a principal 
fiber bundle naturally attached to it, the bundle of linear frames. 
The set of linear frames at a point of the manifold is isomorphic to 
the real linear group $GL(m, R)$ of real $m \times m$ matrices (with 
$m = dim M$), so that the natural bundle is a principal fiber bundle 
with this group. As we are speaking of Minkowski spacetime, whose 
main characteristic is the Lorentz metric, a special role will be 
reserved to the sub-bundle of pseudo-orthogonal frames (that is, of 
the frames which are orthogonal according to the Lorentz metric). 

Let us be a bit more precise. A linear frame at point $p$ of the base 
space is chosen as follows. Take the tangent space at $p$, $T_pM$. It 
is isomorphic to the Euclidean space $E^m$. On $E^m$ there is a 
canonical frame, formed by the unit one-dimensional vectors $\delta_k$, 
having $1$ at the $k$-th entry and zero everywhere else. We choose a 
basis, or frame, by transplanting this one to $T_pM$. Formally, a 
linear frame $\{b_k\}$ is given by a mapping $b: E^m \rightarrow T_pM$, 
$b(\delta_k) = b_k$. Thus, different frames correspond to different 
choices $b$ of which $b_k$ correspond to the unit vector $\delta_k$. 
In strict relation to these vectors there is a canonical basis for the 
group Lie algebra, given by "unit" matrices $\Delta^{\alpha}{}_{\beta}$. 
These matrices have $1$ at the $\alpha-\beta$ entry and zero everywhere 
else. Now, given any metric $\eta$ (here the Lorentz metric), we define 
matrices $J_{\alpha \beta} = \eta_{\alpha \gamma} \Delta^{\gamma}{}_
{\beta} - \eta_{\beta \gamma} \Delta^{\gamma}{}_{\alpha}$. Then the $J_
{\alpha \beta}$'s generate the Lie algebra of the orthogonal group of 
$\eta$ (notice that from this notation comes the use of double indices 
for spacetime geometrical objects, at a difference with geometrical 
objects related to other bundles). In our case, the $J_{\alpha \beta}$'s 
will generate the Lorentz group. The frame vectors are then taken to be 
orthogonal, and a sub-bundle results, the bundle of Lorentzian frames. 
A linear connection will 
be the $1$-form $\Gamma = \Delta^{\alpha}{}_{\beta} \Gamma_{\alpha}
{}^{\beta} = \Delta^{\alpha}{}_{\beta} \Gamma_{\alpha}{}^{\beta}{}_
{\mu} dx^{\mu}$. A connection defines a covariant derivative, which 
acts on any object in a well-defined way. A Lorentz connection will 
be  $\Gamma = J_{\alpha \beta} \Gamma^{\alpha \beta} = J_{\alpha 
\beta} \Gamma^{\alpha \beta}{}_{\mu} dx^{\mu}$, and its curvature, 
which is its own covariant derivative, will be $F = \frac{1}{2} 
J_{\alpha \beta} F^{\alpha \beta} = \frac{1}{2} J_{\alpha \beta} 
F^{\alpha \beta}{}_{\mu} dx^{\mu}$ with
$$
F^{\alpha \beta}{}_{\mu \nu} = \partial_{\mu} \Gamma^{\alpha \beta}{}_
{\nu} - \partial_{\nu} \Gamma^{\alpha \beta}{}_{\mu} + \Gamma^
{\alpha}{}_{\epsilon \mu} \Gamma^{\epsilon \beta}{}_{\nu} - 
\Gamma^{\alpha}{}_{\epsilon \nu} \Gamma^{\epsilon \beta}{}_{\mu} \, . 
$$
From these expressions comes the Bianchi identity,
$$
\partial_{\mu} \tilde{F}^{\alpha \beta \mu \nu} - \Gamma^{\alpha}{}_
{\gamma \mu} \tilde{F}^{\gamma \beta \mu \nu} + \tilde{F}^{\alpha}{}_
{\gamma}{}^{\mu \nu} \Gamma^{\gamma \beta}{}_{\mu} = 0 \, .
$$

This is quite analogous to the usual geometrical background of gauge 
models related to "internal" symmetries, but there is here a deep 
difference. The bundle of linear frames, we have said, is more deeply 
rooted on the base manifold than any other bundle. This is shown by 
the presence of a property which is not present in other bundles. 
The property is called {\it soldering} and is embodied in a special 
$E^m$-valued form on the bundle, the solder form. Being $E^m$-valued 
means that it is of the form $S = \delta_k S^k$, with $S^k$ an usual 
form. It is thus a mapping $T_b GL(m,R) \rightarrow E^m$. It 
establishes a direct relation between tangent spaces of the bundle 
manifold and tangent spaces of the base manifold. Given the 
projection mapping $\pi$ of the bundle and its differential $\pi_*$ 
(which maps tangent vectors), then we have $S = b^{-1} \circ \pi_*$. 
This form is canonical, in the sense that it is always there, for 
any differentiable manifold. Now, choosing a frame is always done 
by a section, a mapping $\sigma: M \rightarrow$ bundle. Differential 
forms on the bundle are brought back to $M$ by the pull-back 
$\sigma^*$ of $\sigma$, the dual of its differential. Each Lorentz 
frame (tetrad, vierbein, fourleg) is given by $\sigma: M \rightarrow$ 
bundle, $\sigma: p \rightarrow \{h^{\alpha}(p)\}$, with $\pi \circ 
\sigma(p) = p$. Now, it so happens that, if $\sigma$ chooses the 
frame $h^a$, then its pull-back of the solder form gives back 
precisely $h^a: \sigma^*(S^{\alpha}) = h^{\alpha} = h^{\alpha}{}_
{\mu} dx^{\mu}$. The presence of the solder form allows one, if 
given a metric $\eta$ on $R^m$, to "transform" it into a metric $g$ 
on the base manifold $M$, by $g(X,Y) = \eta(b^{-1} X, b^{-1} Y)$. 
This is the deep reason for the usual way of writing a metric in 
terms of the tetrads, as the last expression is just $g_{\mu \nu} = 
\eta_{\alpha \beta} h^{\alpha}{}_{\mu} h^{\beta}{}_{\nu}$. 

The presence of the solder form has an important consequence: given 
a connection, there exists another natural characteristic of it, 
besides the curvature. It is its torsion, which is the covariant 
derivative of the solder form. As the latter is expressed on $M$ by 
the tetrad, the torsion appears as the  covariant derivative of the 
tetrad fields,
$$ 
T^{\alpha}{}_{\mu \nu} = \partial_{\mu} h^{\alpha}{}_{\nu} - 
\partial_{\nu} h^{\alpha}{}_{\mu} + \Gamma^{\alpha}{}_{\epsilon 
\mu} h^{\epsilon}{}_{\nu} - \Gamma^{\alpha}{}_{\epsilon \nu} 
h^{\epsilon}{}_{\mu} \, .
$$ 
And it turns out that also an extra Bianchi identity holds,
$$
\partial_{\mu} \tilde{T}^{\alpha \mu \nu} - \Gamma^{\alpha}{}_
{\gamma \mu} \tilde{T}^{\gamma \mu \nu} + \tilde{F}^{\alpha}{}_
{\gamma}{}^{\mu \nu} h^{\gamma}{}_{\mu} = 0 \, .
$$
The Bianchi identity is half the field equations of a gauge theory 
(for example, the first pair of Maxwell's equations). We are thus to 
expect that, if we build up a gauge model related to the bundle of 
frames, we have some extra field equations. Thus, the special 
"tight-bound" relation of the linear bundle to the base manifold 
engenders torsion, and this is submitted to an extra Bianchi 
identity. Torsion is nevertheless quite absent in other bundles, 
like those related to usual (internal) gauge theories. Notice that 
"absent' is quite different from "null". The fact that $T = 0$ has 
deep consequences in geometry (that is the general definition of a 
Riemannian space, one with vanishing torsion). It remains for us that 
the presence of $T$ will, already from the start, establish a 
difference for any gauge theory related to the geometry of spacetime. 

Another step is the following: the space ${\bf R}^4$, endowed with 
the Lorentz metric, becomes ${\bf E}^{3,1}$ and can be identified to 
the translation group ${\bf T}^{3,1}$. The whole thing can be then 
rewritten in terms of the bundle of the affine linear bundles. Once 
reduced by the imposition of Lorentz-orthogonality, the bundle is a 
principal bundle with the Poincar\'e group as the structure group 
(gauge group). One might call it the bundle of the affine orthogonal 
frames.
	
Define then the gauge potentials $\Gamma = J_{\alpha}{}^{\beta} 
\Gamma^{\alpha}{}_{\beta \mu} dx^{\mu}$ for the Lorentz sector, and 
$B = T_{\alpha} B^{\alpha}{}_{\mu} dx^{\mu}$ for the translational 
sector. Accordingly, an extra field strength will appear: 
$$\tau^{\alpha}{}_{\mu \nu} = \partial_{\mu} B^{\alpha}{}_{\nu} - 
\partial_{\nu} B^{\alpha}{}_{\mu} + \Gamma^{\alpha}{}_{\epsilon \mu} 
B^{\epsilon}{}_{\nu} - \Gamma^{\alpha}{}_{\epsilon \nu} B^{\epsilon}{}_
{\mu} \, .
$$ 
Given then the Poincar\'e group, with its structure constants, one 
writes directly the vacuum Yang-Mills equations: 
$$
\partial_{\mu} {F}^{\alpha \beta \mu \nu} - \Gamma^{\alpha}{}_{\gamma 
\mu} {F}^{\gamma \beta \mu \nu} + {F}^{\alpha}{}_{\gamma}{}^{\mu \nu} 
\Gamma^{\gamma \beta}{}_{\mu} = 0 \, ;
$$
$$
\partial_{\mu} {\tau}^{\alpha \mu \nu} - \Gamma^{\alpha}{}_{\gamma \mu} 
{\tau}^{\gamma \mu \nu} + {F}^{\alpha}{}_{\gamma}{}^{\mu \nu} B^{\gamma}{}_
{\mu} = 0 \, .
$$ 
Because translation generators have dimensions, the corresponding fields 
have rather strange dimensions themselves: $[B] = 0$ and $[\tau] = 1$. 
The relation between the translational gauge potential and the tetrads is 
given by~\cite{24}:
$$ 
h^{\alpha}{}_{\mu} = \frac{\partial x^{\alpha}}{\partial x^{\mu}} + 
k B^{\alpha}{}_{\mu} \, .
$$
Thus, $B$ is the non-trivial, anholonomous part of the tetrad. There 
is more here than a mere coordinate transformation. The torsion relates 
to the field strengths $F$ and $\tau$ by  $T = \tau - F x$. The Yang-
Mills equations become 
\be
\partial_{\mu} {F}^{\alpha \beta \mu \nu} - \Gamma^{\alpha}{}_{\gamma 
\mu} {F}^{\gamma \beta \mu \nu} + {F}^{\alpha}{}_{\gamma}{}^{\mu \nu} 
\Gamma^{\gamma \beta}{}_{\mu} = 0 \, ;
\label{1} 
\ee
\be
\partial_{\mu} {T}^{\alpha \mu \nu} - \Gamma^{\alpha}{}_{\gamma \mu} 
{T}^{\gamma \mu \nu} + {F}^{\alpha}{}_{\gamma}{}^{\mu \nu} h^{\gamma}
{}_{\mu} = 0 \, .
\label{2}
\ee 
It is remarkable that the above equations are just the Bianchi 
identities written for the dual fields, so that duality symmetry is 
respected. 

These equations would give an answer to the issue referred to in the 
introduction concerning the complete treatment of the local spacetime 
symmetries. The Noether current for the translational invariance appears 
as the source in the torsion equation, whereas the Noether current for 
the rotational and boost invariance, the relativistic angular momentum 
density, comes up as the source for the curvature equation. The presence 
of a dynamical equation for the torsion gives the theory an advantage 
over the theories of Einstein-Cartan type. Equations (1) and (2) have 
been proposed directly by Popov and Daikhin~\cite{pop}, but for them 
the tetrads, and not their non-trivial parts were supposed to be the 
translational gauge potentials. There are great difficulties~\cite{25} 
with this interpretation. The tetrads, as we have seen, are always there. 
There is no way of anullating them, there would be no way to describe 
the absence of gravitational field. The theory has some very good points: \\
(i) in the sourceless static spherically symmetric case, one finds 
Newton's law with  $L = \sqrt{4 \pi G}$; \\ 
(ii) if we put by hand $T = 0$, the second equation gives ${F}^{\alpha}
{}_{\gamma}{}^{\mu \nu} h^{\gamma}{}_{\mu} = 0$, which is the vacuum 
Einstein's equation; \\
(iii) technically, as gauge fields have spin 1, particle-particle 
interaction has opposite sign to particle-antiparticle interaction; 
but, the indices $\alpha, \mu$, etc in $B^{\alpha}_{\mu}$ are all 
related to spacetime, so that we can possibly find a way of taking 
$B$ as a spin 2 field~\cite{26}.

There are, however, differences with respect to usual gauge theories. 
For example, the Poincar\'e group is non-semisimple, which implies 
that there is no Lagrangian yielding its Yang-Mills equations. 
Moreover, as it acts on spacetime, it has kinematic representations, 
which will respond for the universality of gravitation. And finally, 
it is worth mentioning that the fields have abnormal dimensionalities: 
$[B] = [h] = 0$ and $[\tau] = [T] = 1$. 

\section{Quantization: from Poincar\'e to de Sitter}

Well, the goal is to obtain a quantum theory! Thus, we take the equations 
and try to quantize them. We would hope to be able to use all the so 
well lubricated machinery of gauge theories: partition function, 
Faddeev-Popov, etc. We start by looking for the Lagrangian density. 
And then, we have the first shock: the above Yang-Mills equations have 
no Lagrangian density, they do not come from any action through an 
extremal principle~\cite{10a}. This is a negative point, but it should 
be remembered that at least one basic equation of classical Physics 
is in the same case, the Euler (and the Navier-Stokes) equation of 
fluid dynamics. 

Whether or not an equation or set of equations come from an action 
functional is the object of the Helmholtz-Vainberg theorem, which can 
be put in a very simple form if we generalize exterior differential 
calculus to functionals~\cite{28}. We can show that, for the Yang-Mills 
equation to have a Lagrangian, the gauge group must be in one of the 
following cases: (i) a semisimple group; (ii) an abelian group; (iii) a 
direct product of an abelian and a semisimple group. As the Poincar\'e 
group is in neither of these cases, the equations have no Lagrangian. 
We are thus condemned to try to quantize the theory without resource to 
any of the usual Lagrangian-based techniques. It is possible in principle 
to quantize a theory directly from the field equations, as in the 
formalism of K\"all\'en-Yang-Feldman~\cite{29}. One expects difficulties, 
of course. For example, it will be impossible to use the Faddeev-Popov 
trick, so that ghosts should be introduced in the rather non-systematic 
way of Feynman's "polish paper"~\cite{29a}. Actually, one does not even 
have to face these problems --- a second shock is waiting for us before 
that: the theory is not quantizable! It presents a sickness, which we 
can call vertex inconsistency, or perturbative incoherence. There exists 
a general procedure~\cite{30} to check whether or not a set of field 
equations leads to a coherent theory (as a curiosity, if you try to take 
the Navier-Stokes equation as a model field theory equation, it is 
incoherent: it leads to a theory which cannot be quantized). This 
general formalism shows that every Lagrangian theory is automatically 
consistent, but that there are in principle also non-Lagrangian theories 
which are consistent. But if we apply it to Yang-Mills equations for 
non-semisimple groups, we find that they are never consistent. Well, 
what interests us here is the bad result: the Yang-Mills equations for the 
Poincar\'e group cannot be quantized.

We should not, however, minimize the arguments favouring a Poincar\'e 
group gauge theory for gravitation. The group is {\it the} classical 
group for relativistic kinematics, it classifies all elementary 
particles, etc. On the other hand, one sees that such a theory is 
essentially classical, it cannot be given a quantum version. Is there 
a way out of the above difficulty? Actually, there is: the same general 
procedure which tells whether a set of field equations leads or not  
to a coherent theory provides a systematic way to "patch the theory up". 
It allows to obtain the minimum terms which should be added to an 
inconsistent theory in order to make of it a good theory. The method 
is analogous to that of renormalization: one adds extra terms so that 
the overall theory becomes tractable. The difference here is that terms 
are to be added to the field equations. 

The simplest form to get a consistent theory is to drop all terms 
coupling $B$ to $\Gamma$ in eqs.(\ref{1}-\ref{2}). The result is a 
consistent Lagrangian theory, which is actually a gauge model for the 
direct product between Lorentz and translation groups. However, as a 
vector field, the gauge potential $B$ should couple to the Lorentz 
gauge potential $\Gamma$. This can be achieved by considering $B$ as 
a source field, and then replacing all ordinary derivatives by 
covariant ones. The result is
\be
\partial_{\mu} F^{\alpha \beta \mu \nu} - \Gamma^{\alpha}{}_
{\gamma \mu} F^{\gamma \beta \mu \nu} + F^{\alpha}{}_{\gamma}
{}^{\mu \nu} \Gamma^{\gamma \beta}{}_{\mu} = L^{-2} \tau^{\alpha 
\mu \nu} B^{\beta}{}_{\mu}
\label{3}
\ee
\be
\partial_{\mu} \tau^{\alpha \mu \nu} = \Gamma^{\alpha}{}_
{\gamma \mu} \tau^{\gamma \mu \nu} \, .
\label{4}
\ee
The factor $L^{-2}$ has to be introduced for dimensional reasons. 
These equations can be derivable from the Lagrangian ${\cal L} = 
- \frac{1}{4} (F^2 + L^{-2} \tau^{2})$, from which we can see that 
the sources appearing in eqs.(\ref{3}-\ref{4}) are respectively 
the spin density and the spin energy-momentum tensor of the $B$ 
field.

The next step is to find a way of introducing the term $F^{\alpha}
{}_{\gamma}{}^{\mu \nu} B^{\gamma}{}_{\mu}$ into eq.(\ref{4}) in 
order to go as close as possible to the eqs.(\ref{1}-\ref{2}) of 
the Poincar\'e gauge model. This can be achieved by adding the 
term $- (1/2L^2)F_{\alpha \beta}{}^{\mu \nu} B^{\alpha}{}_{\mu} 
B^{\beta}{}_{\nu}$ to the Lagrangian, which will contribute 
correctly to give eq.(\ref{2}), as well as will give a further 
contribution to eq.(\ref{1}). The new Lagrangian becomes
\be
{\cal L} = - \frac{1}{4} F_{\alpha \beta}{}^{\mu \nu} \left(
F^{\alpha \beta}{}_{\mu \nu} + 2 L^{-2} B^{\alpha}{}_{\mu} 
B^{\beta}{}_{\nu} \right) - 4 L^{-2} \tau_{\alpha}{}^{\mu \nu} 
\tau^{\alpha}{}_{\mu \nu} \, ,
\label{5}
\ee 
which represents a rather complicated, but consistent theory.

Once we arrive at this point, we can examine the renormalization 
question. The situation is found to be quite analogous to the meson 
electrodynamics case: as it stands, the theory is nonrenormalizable, 
but it so happens that a counterterm of quartic type, $(- 1/4 L^{4}) 
(B^{\alpha}{}_{\mu} B_{\alpha}{}^{\mu} B^{\beta}{}_{\nu} B_{\beta}
{}^{\nu})$ added to the Lagrangian (\ref{5}) makes the theory 
renormalizable. The Lagrangian then becomes
\be 
{\cal L} = - \frac{1}{4} \left[ \left(F + 2 L^{-2} B B \right)^2 - 
4L^{-2} \tau^2 \right] \, .
\label{6}
\ee
And here comes the main point. We have said that only a few theories 
are perturbatively renormalizable, and we have arrived at one of them. 
It should be of one of the types given previously. Once we think in 
that way, it comes not as a surprise that the above theory --- which 
is a Poincar\'e gauge theory corrected so as to be quantizable, and 
corrected so as to be renormalizable --- is also a gauge theory. In 
effect, if we redefine the fields by absorbing the abnormal dimensions: 
$\Gamma^{\alpha 5}{}_{\mu} := L^{-1} B^{\alpha}{}_{\mu}$, 
$F^{\alpha 5}{}_{ \mu \nu} := L^{-1} \tau^{\alpha}{}_{\mu \nu}$, 
and put $F^{\alpha \beta}{}_{\mu \nu} := F^{\alpha \beta}
{}_{\mu \nu}(old) + L^{-2}(B^{\alpha}{}_{\mu} B^{\beta}{}_{\nu} - 
B^{\alpha}{}_{\nu} B^{\beta}{}_{\mu})$, we find that the above 
Lagrangian is that of a gauge theory for semi-simple group, 
actually a de Sitter group! In this way, the de Sitter gauge 
theory comes up as the quantizable, renormalizable Poincar\'e 
gauge theory~\cite{31}.

\section{Final comments}

There are, of course, some problems. One of them concerns the 
non-compact character of the de Sitter group, and to the consequent 
unboundedness of the Hamiltonian. This problem is plausibly solved 
through the choice of convenient boundary conditions~\cite{fro}. A 
second problem concerns interpretation of the length parameter $L$. 
Instead of the usual geometrical picture of a Minkowski space 
tangent to each point of  spacetime, the tangent spaces are now 
de Sitter spaces. In this case, $L$ is a length parameter attached 
to any de Sitter space, its pseudo-radius. The usual picture is 
here reversed: spacetime itself is Minkowski space, while the 
tangent spaces are the curved de Sitter spaces. In the case of 
DS(3,2), the covering space is topologically $E^4$ and the tangent 
spaces are at least diffeomorphic to a flat space.

In order to accept a de Sitter group as the group of relativistic 
kinematics in replacement to the Poincar\'e group, one should accept 
that it, and not Poincar\'e, classifies the elementary particles. 
This is acceptable for present-day knowledge, provided $L$ is large 
enough. The contact with usual Minkowski results could then be 
obtained through the use of stereographic coordinates. The definitions 
of energy and momentum would be changed, though in each case the 
departures from their usual meanings would be of order (at least) 
$L^{-1}$.

The de Sitter approach has many fine points, coming mainly from 
formal aspects~\cite{32}. One of them, noticed by Dirac a long 
time ago, refers to the $\gamma$ matrices. The de Sitter group 
puts $\gamma_5$ on an equal footing with other $\gamma$'s, as 
its generators in the spinor representation are $\sigma_{\mu \nu} = 
\frac{i}{2}[\gamma_\mu, \gamma_\nu]$ and $\sigma_{5 \nu} = 
\frac{i}{2}[\gamma_5, \gamma_\nu]$. This means that chiral symmetry 
has a foot in the theory and suggests a partial symmetry breaking. 
Another one is purely geometrical: the de Sitter groups coincide 
with the bundles of Lorentzian frames on a de Sitter space: DS(3,2) = 
SO(3,2)/SO(3,1) and DS(4,1) = SO(4,1)/SO(3,1). 

There are prospective effects concerning both cosmology and parti\-cle 
phys\-ics. Let us only touch two of them. \\
1. We have said that the Poincar\'e theory leads to Newton's law in 
the appropriate limit. From the additional terms appearing in the 
Yang-Mills equations of the de Sitter theory, we should expect some 
modification of this law at distances of the order of $L$. Changing 
Newton's law would alter the interpretation of data concerning the 
missing mass problem in galaxies and clusters. This would suggest 
values of cosmological scale for the universal constant $L$. \\
2. If one of the de Sitter groups is to replace the Poincar\'e group 
as the basic, kinematic classifying group, a lot is changed. The concept 
of energy, as said, must be reformulated. There is more: both CP and CPT 
will be universally violated in their present-day formulation. By 
"universally violated" we mean that all interactions will feel the 
same effects, so that in CP (always in its usual formulation) there 
will be no reason to privilege the $K^0 {\bar K^0}$ system. Also the 
"new CPT" will differ from the usual operation.
Nevertheless, the violation should be less than something, as no 
universal violation has been observed. Since the differences are 
proportional to factors of $L^{-1}$, $L$ can always be taken large 
enough to make them negligible.
Or, if experiments come to detect such an effect, they will also fix 
the value of $L$. The $K^0 {\bar K^0}$ system, which allows very 
high precision measurements of the mass splitting, seems to be a 
good candidate.

\vspace{0.5 cm}
\section*{Acknowledgments}

The authors would like to thank CNPq, Brazil, for partial financial support.

\vfill \eject

\end{document}